\documentclass{aa} 
\usepackage{graphicx,natbib}

\begin{document}
 
\title{3-D ionization structure (in stereoscopic view) of Planetary 
Nebulae: the case of NGC 1501}

\titlerunning{3-D ionization structure of NGC 1501}

\author{ R. Ragazzoni\inst{1}  \and E. Cappellaro\inst{1} \and 
S. Benetti\inst{1}
\and M. Turatto\inst{1} \and F. Sabbadin\inst{1}}

\offprints{F. Sabbadin}

\institute{Osservatorio Astronomico di Padova, vicolo dell'Osservatorio 5,
I-35122 Padova, Italy}

\date{Received 9 January 2001; accepted 29 January 2001}

\abstract {Long-slit echellograms of the high excitation planetary nebula NGC
1501, reduced according to the methodology developed by Sabbadin et
al. (2000a, b), allowed us to obtain the ``true'' distribution of the
ionized gas in the eight nebular slices covered by the spectroscopic
slit.
A 3-D rendering procedure is described and applied, which assembles 
the 
tomographic maps and rebuilds the spatial structure.
The images of NGC 1501, as seen in 12 directions separated by 
15$^o$, 
form a series of stereoscopic pairs giving surprising 3-D views in as 
many directions.
The main nebula consists of an almost oblate ellipsoid of moderate 
ellipticity (a$\simeq$44 arcsec, a/b$\simeq$1.02, a/c$\simeq$1.11), 
brighter in the equatorial 
belt, deformed by several bumps, and embedded in a quite homogeneous, 
inwards extended cocoon.
Some reliability tests are applied to the rebuilt nebula; the radial 
matter profile, the 
small scale density fluctuations and the 2-D (morphology) - 3-D (structure) 
correlation are presented and analysed. 
The wide 
applications of the 3-D reconstruction to the 
morphology, physical conditions, ionization parameters and evolutionary status 
of expanding nebulae in general (planetary nebulae, nova and 
supernova remnants, shells around Population I Wolf-Rayet stars, nebulae 
ejected by symbiotic stars, bubbles surrounding early spectral type main 
sequence stars etc.) are introduced.
}
\maketitle

\section{Introduction}

The planetary nebula (PN) phase (common to most of intermediate and
low mass stars in their late evolution; Iben, 1984; Vassiliadis \& 
Wood, 1993; Sch\"onberner, 1997)
is characterized by an extreme variety of morphologies: disks,
butterflies, point-symmetric or multiple shells, hourglasses etc.
\citep{greig,balick,schwarz,manchado}.

Different mechanisms, acting before, during and after the nebula
ejection, can contribute to generating the observed shapes. These
include the mass-loss history, stellar rotation and/or precession,
magnetic fields, duplicity of the central star, wind interactions,
hydrodynamical processes, ionization and recombination
\citep{tylenda,morris,pascoli92,dwar,garcia}.

In order to disentangle the physical processes forming and shaping a
PN, the first - and most difficult - observational step is the
nebula de-projection, i. e. the recovery of the ionized gas 3-D structure. 
In this paper, which is the result of
a long-standing effort (Sabbadin, 1984, Sabbadin et al., 1985, 1987,
2000a, 2000b), we obtain the spatial distribution of the PN matter
by means of high resolution spectroscopy.

As a test case we selected NGC~1501 for which we have already
presented a partial tomographic analysis (Sabbadin et al. 2000b,
hereafter P1). 
In P1, long-slit echellograms at four position angles gave the 
radial density profile, the mean electron temperature and 
turbulence and a rough 3-D model. The accurate spatial reconstruction 
was deferred to a more complete spectroscopic coverage.

Recently we have observed NGC 1501 at four additional position 
angles and developed a rendering procedure which integrates the 
tomographic slices and rebuilds the detailed spatial distribution of 
the ionized gas.

This work, in many respect the logical continuation of
P1, is structured as follows: the nebula and the observational
material are introduced in Sect. 2; in Sect. 3 we illustrate the
reduction procedure, in particular, how to derive the electron density
profile from the H$\alpha$ flux of the zero-velocity pixel 
column; Sect. 4
contains the tomographic maps and Sect. 5 describes and analyses
the spatial model; a general discussion is presented in Sect. 6 and
conclusions are drawn in Sect. 7.

\section{The nebula and the observational material}

\begin{figure}
%\resizebox{\hsize}{!}{\includegraphics*{MS1018f1.eps}}
\caption{Spectroscopic slit positions and position angles, 
superimposed on a broad-band R frame of NGC 1501 taken with the 3.58m
Italian National Telescope (TNG).  The seeing was FWHM=0.76
arcsec. North is up and East is to the left.}
\label{figura1}
\end{figure}

The high excitation PN NGC 1501 (PNG144.5+06.5, Acker et al., 1992) is
``a mass of irregular condensations, with some big holes''
\citep{minkowski} ``bearing a resemblance to the convolutions of
the brain'' \citep{pease}.

The literature-based data for the whole system (nebula + exciting star) are
summarized in Table 1.

\begin{centering}
\begin{table*}
\caption{Basic data of NGC 1501 taken from the literature}
\begin{tabular}{ll}
\hline
\\
\multicolumn{2}{c}{\bf{NEBULA}}\\
coordinates (2000.0)& $\alpha$: 04$^h$ 06$^m$ 59.7$^s$ $    \delta$: +60$^o$ 55' 14''\\
apparent size [arcsec]& 56x48 (Curtis, 1918); 68 faint, roundish halo (P1)\\
log F(H$_{\beta}$) [mW$\times$m$^{-2}$]& $-11.28$ (Collins et al., 1961); $-11.20$ 
(Acker et al., 1991)\\
c(H$\beta$)& 0.96 (Kaler, 1976); 1.1 (Stanghellini et al., 1994); 1.11 (Ciardullo et 
al., 1999); 1.05 (P1)\\
excitation class& 8 peculiar (Page, 1942)\\
individual distance [Kpc]& 2.0 (Acker, 1978); 1.4 (Pottasch, 1983); 1.2 (Sabbadin, 1986)\\
& (all based on the average extinction in the galactic disk)\\
statistical distance [Kpc]& 1.45 (O'Dell, 1962); 1.20 (Cahn 
\& Kaler, 1971); 1.78 (Cudworth, 1974); 1.16 (Cahn, 1976); \\
&1.16 (Acker, 1978); 1.10 (Daub, 1982); 1.10 (Maciel, 1984); 
 0.90 (Amnuel et al., 1984); \\
& 1.50 (Sabbadin, 1986); 1.16 (Cahn et al., 1992); 1.31 (Van de Steene \& Zijlstra, 
1994); \\
& 1.21-1.75 (Zhang, 1995)\\
adopted distance [Kpc]& 1.30 \\
mean linear radius [pc]& 0.164 \\
expansion velocity [km s$^{-1}$]& 39 in [OIII] (Robinson et al., 
1982); 38 in [OIII] and HI (Sabbadin \& Hamzaoglu, 1982);\\
& 40 in HeII, [OIII], HI and [NII] (Neiner et al., 2000); 38 to 
55 in [OIII] and HI (P1)\\
turbulence [km s$^{-1}$]& 10 (Neiner et al., 2000); 18 (P1)\\
structure& prolate spheroid of moderate ellipticity (Sabbadin 
\& Hamzaoglu, 1982);\\ 
& thin ellipsoid deformed by a pair 
of large lobes along both the major and the intermediate \\
& axes and by a heap of bumps spread over the whole 
nebular surface (P1)\\
\\
\multicolumn{2}{c}{\bf{CENTRAL STAR}} \\
m$_V$&  14.23 (Tylenda et al., 1991); 14.36 (Ciardullo et al., 1999)\\
spectrum&  WC6 (Swings \& Swensson, 1942); WC-OVI (Aller, 1976); WC4 (Tylenda et al., 
1993)\\
logT$_{star}$& 4.98 (TZ$_{HeII}$; Sabbadin, 1986); 4.91 (TZ$_{HeII}; $ 
Stanghellini et al., 1994);\\
& 5.13 (atmosphere model; Koesterke \& Hamann, 1997)\\
mass loss rate [M$_\odot$yr$^{-1}$]& 5.2x10$^{-7}$ (Koesterke \& 
Hamann, 1997)\\
terminal wind velocities [km s$^{-1}$]& 1800 (Koesterke \& 
Hamann, 1997); 3260 to 3460 (Feibelman, 1998)\\
note& the star is a nonradial g-mode pulsator with periods 
ranging from 5200s to 1154s \\
& (Bond et al., 1998; Ciardullo \& Bond, 1998)\\
\\
\multicolumn{2}{l}{Further informations on NGC 1501 are contained in 
 Pottasch (1983) and Acker et al. (1992).} \\
\hline
\end{tabular}
\end{table*}
\end{centering}

$\lambda$$\lambda$4500-8000 \AA, spatially resolved spectra of NGC
1501 (including flat field, Th-Ar calibration and spectrophotometric 
standard)
were secured with the Echelle spectrograph attached to the Cassegrain
focus of the 1.82m Asiago (Italy) telescope and equipped with a Thompson
1024x1024 pixels CCD (spatial scale=0.79 arcsec pix$^{-1}$; spectral
scale=9.0 km s$^{-1}$ pix$^{-1}$).

Eight position angles were covered in two observing runs (the slit
positions are shown in Fig.~1 superimposed on a broad-band R frame of
the nebula):

Run 1) P.A. = 10$^o$, 55$^o$, 100$^o$ and 145$^o$ on December 1998 (non
photometric nights; seeing 1.4 - 1.7 arcsec; spectral resolution 13.5
km s$^{-1}$; spectra used in P1);

Run 2) P.A. = 30$^o$, 80$^o$, 125$^o$ and 170$^o$ on 
December 1999 (photometric nights; seeing 1.2 - 1.4 arcsec; spectral
resolution 12.0 km s$^{-1}$).

In all cases the exposure time was 1800 seconds.

A detailed description of the spectroscopic characteristics of NGC 
1501 is given in P1. In short: it is a high ionization, density bounded 
 PN; stratification effects are negligible; the position-velocity 
maps can be fitted by a triaxial ellipsoid deformed by several 
hemispheric bubbles protruding from the central figure; the main 
emission occurs in a narrow shell whose density profile is steep 
outwards and flatter inwards.

We believe that a general observational note is appropriate here, 
concerning the common employment by most authors of an interference 
filter when obtaining
echelle spectra of PNe (to isolate a single order
containing one or more interesting emissions, like H$\alpha$ + [NII],
and to avoid the superposition of orders). No filter was used in our 
observations, for the following reasons:

\begin{description}
\item[-] PNe present a discrete spectrum and the superposition
of lines belonging to different orders is quite improbable (even with
a slit covering several orders);

\item[-] the introduction of an interference filter, reducing the spectral
range from several hundred \AA\/ to a few dozen \AA\/, drastically
limits the number of usable emissions per frame;

\item[-] a careful reduction procedure allows us to take into account 
all the possible sources of
inaccuracy (line curvature, order tilting, optical distorsions,
spectral sensitivity, CCD cosmetics) over the whole frame, thus
obtaining the flux and wavelength calibrated, spatially resolved
structure of all emissions present in the echellogram (some dozens in
the case of the brightest PNe, representing the main ionic
species).
\end{description}

\section{Reduction procedure: the radial electron density ($Ne$) profile from 
the H$\alpha$ absolute flux in the zero-velocity pixel column}

The spectral reduction process was the same as described in P1, but for
the absolute calibration of the radial $Ne$ profile. Due to night 
sky variability, in P1 we were forced to obtain the electron 
density of NGC 1501 by comparison with NGC 40 (both nebulae observed 
on the same nights and with the same instrumental setup). This 
was unnecessary for ``run 2'', since the sky 
was photometric and the proper nebular flux calibration was
performed using the spectrophotometric standard.  

Thus, the starting point is the classical expression

\begin{equation}
4\pi D^2 F(H\alpha)= h \nu_{3,2} \int_{0}^{R}\alpha_{3,2}\,
N(H^+)Ne\,\epsilon 4
\pi r^2 dr
\end{equation}

\noindent
where: 

D is the distance, $F(H\alpha)$ is the H$\alpha$ flux corrected 
for extinction, $\nu_{3,2}$ is the H$\alpha$ frequency, $\alpha_{3,2}$ 
is the H$\alpha$ recombination coefficient, $N(H^+)$ is the ionized 
hydrogen density and $\epsilon$ is the filling factor, representing 
the fraction of the nebular volume which is filled by matter at density $Ne$.

We restrict the integration to the portion of the nebula
identified by the zero-velocity pixel column (corresponding to the
gas which is expanding perpendicularly to the line of sight).

\begin{figure}
%\resizebox{\hsize}{!}{\includegraphics*{MS1018f2.eps}}
\caption{Radial electron density profiles at the four position angles of 
``run 2'', obtained from the absolute flux in the H$\alpha$ zero-velocity 
pixel columns.} 
\label{figura2}
\end{figure}

The H$\alpha$ flux (F$_l$) in each pixel of the zero-velocity column,
corrected for interstellar absorption and instrument resolution (for
details see Sabbadin et al. 2000a), is given by:

\begin{equation}
 4\pi D^2 F_l=4\pi j_{H\alpha} N(H^+) Ne V_l \epsilon_l ~~~~~~~~~~~~{\rm
 [erg\, s^{-1}]}
\end{equation}

\noindent 
where: 

\begin{description}
\item{-} $j_{H\alpha}$ is the H$\alpha$ emission coefficient for the case B
of \citet{baker}; for typical nebular densities its dependence on $Te$
is obtained by interpolation of \citet{brock} data:
 $4\pi j_{H\alpha} = h \nu_{3,2} \alpha_{3,2} = 0.20\times10^{-20} Te^{-0.94}$;

\item{-} $V_l$ is the ``local nebular volume'' defined as
$V_l=A\times\Delta r$, $A$ being the pixel area and $\Delta
r=R\times(\Delta V/V_{\rm exp})$ the radial thickness of the
zero-velocity pixel column;

\item{-} $\epsilon_l$ is the ``local filling factor'', indicating the 
fraction of the local nebular volume, $V_l$, which is actually filled by 
matter with density $Ne$.
\end{description}

The radial electron density profiles of NGC~1501 derived from eq.
(2) at the four position angles of ``run 2'' are shown in Fig.~2
(assuming $Te=11500^\circ$K, $Ne=1.15$ $N({\rm H}^+$), $c({\rm
H}\beta)=1.05$, $D=1.30$ Kpc and, in first approximation,
$\epsilon_l$=1).  Their complex structure confirms (at least
qualitatively) the results already obtained in P1; in particular,
 the main nebula
consists of a thin shell presenting large inwards tails whose extent
is anti-correlated to the height of the density peak.

A quantitative $Ne_{{\rm run 1}} - Ne_{{\rm run 2}}$ comparison can be
performed only at the intersection of the eight slit positions,
i. e. at the apparent stellar position; this needs the accurate
tomographic reconstruction illustrated in the next Section.
 
\section{Tomography}

The tomographic analysis of PNe, originally introduced and applied to
plate spectra by \citet{sabba84} and Sabbadin et al. (1985, 1987),
has been recently extended and refined by Sabbadin et al. (2000a,
b). Its application to the H$\alpha$ emission of ``run 2'' spectra
allows us to derive the electron density maps shown in Figure 3.
Following P1, we adopted a linear distance-velocity relation, an
electron temperature of $11500^\circ$K, a turbulence of 18 km
s$^{-1}$ and a radius of the shell peak at the position of the central
star (in radial direction) of 20 arcsec.

The maps of Fig. 3, combined with the similar maps of Fig. 5 of P1,
give the spatial $Ne$ distribution in the eight nebular slices covered 
by the
spectroscopic slit.  NGC 1501 being a high ionization, density bounded
PN, the $N({\rm H}^+)$ and $N({\rm O}^{++})$ tomographic maps coincide
with the $Ne$ ones (but for a scaling factor of 1.15 and $3.8 \times
10^3$, respectively; see P1).

The comparison of the electron densities in the nebular regions common
to the eight position angles (i. e. at the apparent stellar position)
gives a very satisfactory agreement (to within $\pm$3\%) between
$Ne_{\rm run 1}$ and $Ne_{\rm run 2}$, confirming the validity of the
``escamotage procedure'' adopted in P1.  Our satisfaction is mitigated
by the consideration that, in any case, the inaccuracy in the
absolute $Ne$ calibration based on eq. (2) is of the order of 10\%, 
mainly due to the uncertainty in the distance of the nebula.

The only direct $Ne$ determination in NGC 1501 using diagnostic line
ratios dates from \citet{allerepps}. These authors analysed at low
spectral resolution a small region located at a distance of 20 arcsec
from the central star (in PA$= 125^o$) and derived
$I(6717)/I(6731)=0.78$, corresponding to $Ne= 1200\,{\rm cm}^{-3}$
(for $Te=11500^\circ$K). From the tomographic map at PA$=125^o$
(Fig. 3) we obtain, at a distance of 20 arcsec in the E-SE sector,
density peaks of 950-970 cm$^{-3}$, i.e. our $Ne(SB)$ ($SB$ stands for 
``surface
brightness'') is 20\% lower than the corresponding [SII] electron density
reported by
\citet{allerepps}.

\begin{figure*}
%\resizebox{18cm}{!}{\includegraphics*{MS1018f3.eps}}
\caption{True electron density structure in the four slices of NGC 1501 
covered by the spectroscopic slit during ``run 2''. The lowest density
shown is $Ne=200$ cm$^{-3}$, while the highest one ($Ne$ = 1410
cm$^{-3}$) is reached by the approaching knot in PA$= 30^o$
(N-E sector), at an apparent distance of 11 arcsec from the central
star.}
\label{figura3}
\end{figure*}

Possible causes of this discrepancy are (in order of importance):

\begin{description}
\item[a:] small-scale density fluctuations. We adopted a local 
filling factor $\epsilon_l=1$; actually, the surface brightness
method (cf. eq. (2)) gives $Ne\times  \epsilon_l^{1/2}$. On the
other hand, the [SII] line intensity ratio measures the true electron
density of the brightest (i. e. densest) nebular regions; thus the
$Ne(SB)/Ne([{\rm SII}]$) ratio would indicate a local filling
factor $\epsilon_l \simeq 0.66$ in the main shell of NGC 1501;

\item[b:]
distance inaccuracy. From eq. (2) we have that $Ne(SB)\propto
(1/D)^{1/2}$; to obtain $Ne(SB)= Ne([{\rm SII}])$ we need $D_{\rm NGC
1501}=0.83$ Kpc (assuming $\epsilon_l=1$). This value is lower than
any of the estimates reported in the literature (see Table 1);

\item[c:] ionization. NGC~1501 being a high ionization PN, S$^+$ 
is localized in the outer parts of the densest regions, where the
shadowing of the UV stellar flux decreases the gas ionization. Thus, the
H$^+$ and S$^+$ spatial distributions are partially decoupled. The
effects of ionization on the $Ne(SB)/Ne({\rm [SII]})$ ratio are
generally modest; they become important in the case of a radial
density profile which is multi-peaked, very asymmetric and strongly
variable in direction;
  
\item[d:] measurement uncertainty connected to the weakness of low 
ionization 
emissions. A symptom of the difficulties encountered by Aller and Epps
(1976) comes from the
non-detection of the [SII] red doublet at a second position close 
to the nebular edge
(25 arcsec from the nucleus, in PA$=223^\circ$), which is noticeably
brighter than the first one. For this second position (intermediate
between our PA$=30^\circ$ and PA$=55^\circ$) we derive electron
density peaks of $1050 - 1150\,{\rm cm}^{-3}$.  Note that
these $Ne(SB)$ values are only 10\% higher than those obtained for the
first position, indicating that the large flux observed in position 2
is mainly due to projection effects (i.e. thickness of the emitting
layer).
\end{description}

Most likely, to explain the $Ne(SB)-Ne({\rm [SII]})$ discrepancy, we
must look for a suitable combination of all the factors listed above.
New, valuable information will come out of the detailed spatial
reconstruction of the nebular structure, as performed in the next
Section.

\section{3-D spatial reconstruction}
\subsection{Methodology}

Each tomographic map of NGC 1501 can be regarded as an image
$Ne(x,x')$, where $x$ is the distance from the plane of the sky
crossing the central star and $x'$ is measured along the slit.  In
order to recover the spatial electron density distribution
$Ne(x,y,z)$, a suitable reference system is defined: we adopted $x$
pointing toward the Observer, $y$ toward East and $z$ toward North.

A data-cube of a given size $N\times N\times N$ is initialized, 
filled with the raw data and re-mapped into cylindrical coordinates 
$(x,r,\alpha)$, where:

\begin{equation}
r=\sqrt(y^2 + z^2); \; \; \alpha={\rm arctan} \left ( \frac{y}{z} 
\right). 
\end{equation}

Once we have defined the pixel size $p$, we can fill up the $n$ planes 
related to the spectrograms as:

\begin{equation}
Ne(x,r,\alpha)=Ne_i \left( \frac{x}{p} , \frac{x'}{p}
\right).
\end{equation}

Interpolation between the $n$ observed planes is made in a linear fashion 
along the cylindrical coordinates. In other words, the data for 
$\alpha > \alpha_i$ and $\alpha < \alpha_{i+1}$ are given by:

\begin {equation}
Ne(x, r, \alpha) = Ne_i \left(\frac{x}{p},\frac{x'}{p} \right)
\frac{\delta\alpha_i^-}{\Delta\alpha} + Ne_{i+1} \left(\frac{x}{p}, 
\frac{x'}{p} \right)
\frac{\delta\alpha_i^+}{\Delta\alpha}
\end{equation}

\noindent where: 

$\delta$$\alpha_i^-$=$\alpha$ - $\alpha$$_i$,
$\delta$$\alpha$$_i^+$=$\alpha$$_{i+1}$ - $\alpha$ and
$\Delta$$\alpha$=$\alpha$$_{i+1}$ - $\alpha$$_i$.

Finally, one should consider the singularity in $y=z=0$. These points
are treated as an average of the neighbouring pixel-data for
each plane defined by a given $x$.  

In the examples reported here we
used $N=200$, $n=8$ and $p=0.40$ arcsec.

\subsection{Results}

\begin{figure*}
%\resizebox{17.9cm}{!}{\includegraphics*{MS1018f4.eps}}
\caption{Opaque reconstruction of the densest regions of NGC 1501 ($Ne>900 
\, {\rm cm}^{-3}$) as seen from 12 directions separated by 15$^o$.
The line of view is identified by ($\theta$,$\psi$), where $\theta$ is
the zenith angle and $\psi$ the azimuthal angle, representing a
rotation through the first two Euler angles; the upper-right image is
the rebuilt-nebula as seen from the Earth, i. e. from (0,0).  Each
horizontal couple constitutes a ``direct'' stereoscopic pair, allowing
the reader to have 12 3-D views of the nebula in as many directions
(all together covering a straight angle).  Instructions for beginners:
to obtain the three-dimensional vision, look at a distant object and
slowly insert the figure in the field of view (always maintaining your
eyes parallel).  Alternatively, you can use the two small dots in the
upper part of the figure as follows: approach the page
till the two dots merge (they appear out of focus); then recede very
slowly (always maintaining the two dots superimposed) till the image
appears in focus.  A suggestion: be patient. The difficulties
encountered in the beginning soon disappear and you will have
exciting three-dimensional views of the nebula.  }
%\label{figura4}
\end{figure*}

A problem arises when presenting the $Ne$ data-cube of NGC 1501: 
how can we 
render the 3-D on the page? The adopted solution 
is illustrated in the following figures.

 Fig.~4 shows the opaque reconstruction of the densest
nebular regions ($Ne>900\, {\rm cm}^{-3}$) as seen from 12 directions,
separated by $15^\circ$. 

\begin{figure*}
%\resizebox{17.9cm}{!}{\includegraphics*{MS1018f5.eps}}
\caption{Same as Fig. 4, but for the weakest components 
($Ne>300\,{\rm cm}^{-3}$) 
of NGC 1501. To enjoy the stereoscopic view of the nebula in 12
directions, read the instructions contained in the caption of Fig. 4.}
\label{figura5}
\end{figure*}

Fig.~5 is the same as Fig.~4, but for the weakest components 
($Ne>300\, {\rm cm}^{-3}$); at even lower densities the roundish 
halo appears.

Finally, Fig.~6 represents the optical appearance of the rebuilt-nebula 
seen from the same 12 directions of Figs.~4 and 5.

In these figures the line of view is identified by the zenith angle
($\theta$) and the azimuthal angle ($\psi$), corresponding to a
rotation through the first two Euler angles. Thus, for example, the
upper-right image represents the rebuilt nebula as seen from the Earth
(i. e. from (0,0)).  

Let's consider Figs.~4 and 5. Since the line of view of two adjacent images
differs by $15^\circ$, each horizontal couple constitutes a stereoscopic
pair, allowing the reader to enjoy 12 surprising 3-D views of the
nebula in as many directions (instructions are given in the caption of 
Fig.~4). The stereoscopic view is more difficult
in Fig.~6, because of the diffuse, smooth nebular appearance.

Hereafter we will call ``direct stereoscopy'' the 3-D view just 
described; in sub-section 6.3 an alternative method, ``swapped stereoscopy'', 
will be introduced and compared with the ``direct'' one.

For reasons of space, only twelve low resolution nebular structures (at two 
density cuts) are presented, and as many projections. The complete data-cube 
can be directly requested from the authors. Moreover, some movies of NGC 1501 
(+ other pleasantries) will be available soon from a dedicated WEB page.

\begin{figure*}
\begin{centering}
%\resizebox{13.5cm}{!}{\includegraphics*{MS1018f6.eps}}
\caption{Optical appearance of NGC~1501-rebuilt from the same 12 directions of 
Figs. 4 and 5 (and at the same scale). Note that projection
(0,$\psi$)= projection(0,$\psi$$\pm$180$^o$). The upper-right image
corresponds to the rebuilt-nebula as seen from the Earth. In this case
the stereoscopic view is quite difficult, because of the smooth,
diffuse nebular appearance.}
\label{proj}
\end{centering}
\end{figure*}

\subsection{Reliability tests}

The first (and most important) consistency test of our reconstruction
method is based on the comparison of the optical appearance of the true- 
and the rebuilt-nebula, when reduced to the same angular resolution.

To be noticed that the spatial resolution of NGC 1501-rebuilt
is not constant, but anti-correlated to the angular distance from the
central star (this occurs because we observed the nebula at radially
arranged position angles). Thus, taking into account the slit width,
the number of position angles and the nebular angular extension, we
have that the spatial resolution of NGC 1501-rebuilt varies
from 1.5 - 2.0 arcsec (at the central star position) to 3.0 - 3.5
arcsec (at the nebular edge).

The optical appearance of the true nebula under a seeing of 2.0 and 
3.5 arcsec is shown in Fig.~7.  The comparison with NGC 1501-rebuilt 
(Fig.~\ref{proj} upper-right image) is very satisfactory, both
qualitatively and quantitatively. Indeed:

\begin{enumerate}
\item
 all the macro- and micro-characteristics observed in NGC 1501
 (general appearance, holes position, blobs and condensations
 distribution etc.)  are faithfully reproduced by the model;
\item
 the surface brightness profiles of the two nebulae agree to within
 $\pm$10\% along the position angles covered by the spectroscopic slit
 and to within $\pm$20\% at intermediate directions.
\end{enumerate}

Other checks, eg. the comparison of the nebular fluxes at different
radii and/or different positions, the $Ne$ distribution etc.  confirm
that our technique adequately recovers the 3-D nebular
shape.  In conclusion, Figs.  4 to 6 closely represent both the
structure and appearance of NGC 1501 rotating around a N-S axis
centred on the exciting star.

\section{Discussion} 

In general, the foregoing spatial reconstruction performed in
different emissions (i. e. for various ions) allows us to obtain the
detailed 3-D ionization structure of a PN. This is particularly
valuable for objects presenting complex morphologies (butterflies,
poly-polars, multi-shells etc.) and/or large stratification effects
and/or small scale condensations (ansae, knots, FLIERS= fast, low
ionization emitting regions, BRETS= bipolar, rotating, episodic jets
etc.).

Moreover, photo-ionization models at unprecedented resolution of the
whole nebula and of the micro- and macro-structures can be produced by
combining the true spatial structure of the expanding gas with the
ultraviolet stellar flux.

Finally, the 3-D reconstruction can strongly contribute in solving the
PNe ``Problem'', i.e.  the distance: only the detailed knowledge of
both the dynamical and physical nebular properties will give the right
interpretation of the angular expansion measured in first and second
epoch HST and/or radio imagery, thus obtaining precise
expansion-parallax distances (Hajian et al., 1993, 1995; Kawamura \&
Masson, 1996; Terzian, 1997).

From all these points of view NGC 1501 is a complete disappointment: it is
quite faint (i. e. only the brightest emissions can be analysed) and
homogeneous (no knots, jets or ansae) at high excitation and density
bounded (stratification of the radiation is negligible).

In spite of this, the 3-D analysis of our ``ordinary'' PN is important
to introduce the wide potentialities of the spatial reconstruction in
studying at small and large scales both the phenomenology and the
physical processes connected with expanding nebulae (PNe, nova shells,
supernova remnants, nebulae around Population I Wolf-Rayet stars
etc.).
 
\subsection{General 3-D structure of NGC 1501}

\begin{figure}
%\resizebox{\hsize}{!}{\includegraphics*{MS1018f7.eps}}
\caption{Optical appearance of NGC 1501 (the true nebula) under a seeing of 
2.0 arcsec (left) and 3.5 arcsec (right). We have first removed the
stars from the TNG R frame, and later blurred the image by convolution
with a Gaussian having FWHM=2.0 arcsec and 3.5 arcsec, respectively.
The field and the orientation are as in Fig. 1.  You can compare these
images of the true nebula with the one of NGC 1501-rebuilt seen from
(0,0), i.e. with the upper-right image of Fig. 6 (recall that the
spatial resolution of NGC 1501-rebuilt is anti-correlated to the
distance from the central star: 1.5 - 2.0 arcsec at the stellar
position to 3.0 - 3.5 arcsec at the nebular edge).}
%\label{figura7}
\end{figure}

NGC 1501 is a high excitation, density bounded PN, and its spatial
structure is the same in $Ne$, $N({\rm H}^+$) and $N({\rm O}^{++}$),
other than a scaling factor.

Low ionization regions (for example N$^+$, identified by $\lambda$6584
\AA\/ of [NII]) are too weak for a detailed 3-D analysis. Probably
they are localized in the external parts of the densest knots and
condensations, where the shadowing of the ultraviolet stellar flux
lowers the nebular ionization.
 
The results obtained in P1 (the main body of NGC 1501 is a thin
ellipsoid of moderate ellipticity deformed by a pair of large lobes
along both the major and intermediate axes and by a multitude of
smaller bumps spread on the whole nebular surface, making the general
3-D structure like a boiling, tetra-lobed shell) can now be improved and
extended.

The thin, inhomogeneous shell forming the main nebula (Fig. 4) is
characterized by a dense, oblique belt definiing the ``equatorial''
plane, i. e. the plane containing the axes $b$ and $c$ of the
central ellipsoid. The maximum density ($Ne \simeq 1400\,{\rm cm}^{-3}$ for
$\epsilon_l$=1) is reached in the direction of the minor axis.  Only
some arcs and knots belonging to the lobes appear in Fig. 4.

A lumpy, quite homogeneous cocoon at low density completely embeds the
main shell (Fig. 5). The complex radial structure of this envelope
(sharp outwards and broad inwards) cannot be seen here and will be
presented and discussed later (sub-section 6.4).

The ``boiling'' structure of NGC 1501 is clearly visible in Figs. 4 and 5
for the shell's regions with higher spatial resolution, i. e. those
projected on -or close to- the central star when the nebula is seen
from (0,0).

\begin{figure}
%\resizebox{7cm}{!}{\includegraphics*{MS1018f8.eps}}
\caption{Optical appearance of NGC 1501-rebuilt seen from a) (25,75) i. e.
along the major axis of the ellipsoid; b) from ($65,-15$) i. e. along
the intermediate axis and c) from ($-25,-15$) i. e. along the minor
axis. The thin ellipsoid's section and the axes contained in the plane
of each figure are shown.}
\label{ellips}
\end{figure}

To highlight the ``tetra-lobed'' shape, in Fig.~\ref{ellips} we present
the optical appearance of NGC 1501 as seen from:
\begin {description}
\item[a)] ($25,75$), i. e. along the major axis of the ellipsoid; 
\item[b)] ($65,-15$), i. e. along the intermediate axis;
\item[c)] ($-25,-15$), i. e. along the minor axis.
\end{description}

In conclusion: the main body of NGC 1501 is a deformed, almost oblate
ellipsoid of moderate ellipticity ($a \simeq 44$ arcsec, $a/b \simeq 1.02$,
$a/c \simeq 1.11$), denser in the equatorial belt containing the axes
$b$ and $c$. Hydrodynamical processes (possibly Rayleigh-Taylor
instabilities and wind interaction) partially swept up the lower
density regions of the ellipsoid, generating both the hemispheric caps
and the broad inward tail in the radial matter distribution.

The ionized nebular mass ($M_{ion}$, coinciding with the total nebular mass)
is 0.15($\pm$0.03) M$_\odot$ and the dynamical age 5000($\pm$500) years.

As noticed in P1, a small, attached halo extending up to 
34 arcsec from the central star surrounds the main nebula. The roundish, 
homogeneous appearance of the halo suggests that it represents 
photospheric material ejected by the PN progenitor in the late AGB evolution.
Moreover, the radial surface brightness profile, steeper than the r$^{-3}$ 
law expected of a steady-state flow, indicates that the halo corresponds 
to a phase of ``enhanced'' stellar mass-loss. Finally, the electron 
density profile (Fig. 2 of this paper and Fig. 4 of P1) gives a mass-loss 
rate of 3($\pm$1)x10$^{-5}$ M$_\odot$ yr$^{-1}$ and an age of 
1.0($\pm$0.2)x10$^4$ years (assuming $V_{exp}$ = 20 km s$^{-1}$; see 
Habing, 1996 and references therein). 
 
We claim that the halo represents the vestiges of the ``superwind'' 
phase (lasted 5($\pm$1)x10$^3$ years) which generated the nebula. Most of 
the gas was later swept up by wind interaction, and was completely ionized 
by the UV flux of the evolving star, thus producing NGC 1501.
Note that the mean mass-loss rate in the superwind phase resulting 
from the analysis of the main nebula, given by ($M_{ion}$/superwind duration),
coincides with the value just obtained for the halo.
 
At this point a quick comparison with NGC~40 (the first PN
tomographically studied in detail, Sabbadin et al., 2000a) can be
instructive.

NGC 40 is an optically thick, very low excitation barrel-shaped nebula
with thin arcs emerging at both ends of the major axis; it is powered
by a luminous WC8 star presenting a large mass-loss rate.  According
to the ``born-again scenario'' proposed by Iben (1984; see also
Bl\"ocker, 1995 and references therein), the central stars of NGC 40
and NGC 1501 suffered a final thermal pulse in the late post-AGB
phase, ejecting the photospheric strata and exposing the C and O rich
core.

Moreover, both the evolutionary sequence [WC$_{late}$] -
[WC$_{early}$] - [WC-PG1159] - [PG1159] \citep{hamann,koesterke} and
the dynamical ages ($3500\pm500$ yr for NGC 40 and $5000\pm500$ yr
for NGC~1501) would indicate a earlier evolutionary phase for NGC~40.

In $2000\pm500$ yr its size should be comparable with the present size
of NGC~1501, but the structure will be quite different: a dense
(optically thick?) equatorial torus + extended, faint polar caps.

Thus: NGC 40 will most likely evolve into a bipolar (butterfly?) PN, whereas NGC
1501 will keep its ellipsoidal structure, becoming fainter and fainter
(in a few thousand years it will resemble A 43 and NGC 7094, two
extended, filamentary, high ionization PNe excited by hybrid-PG1159
type central stars; see Rauch, 1999 and Feibelman, 2000).
  
The different evolution of the two nebulae can be tentatively
explained by a more massive NGC~40 progenitor, which ejected a higher
nebular mass with a larger density gradient between equatorial and
polar regions, and created a higher mass central star (with respect to
NGC~1501).

Further support for this qualitative scenario comes from a preliminary
analysis of a third ``born-again'' PN at an even earlier evolutionary
phase, BD+303639 (the Campbell's star), observed with TNG + SARG in
nine position angles at spatial and spectral resolutions of 0.7 arcsec
and $\lambda/\Delta\lambda = 115000$, respectively.

\subsection{2-D (morphology) - 3-D (structure) correlation}

Because of its fascinating variety and importance in most aspects of
the advanced evolution of low and intermediate mass stars, the
morphology of PNe has attracted many authors and several efforts have
been made to search for a general classification scheme for the
``morphological forest'' (Greig, 1972; Zuckerman \& Aller, 1986; 
Balick, 1987; Chu et al., 1987; Balick et al., 1992; Stanghellini et al., 
1993; Corradi \& Schwarz, 1995; Manchado et al., 1996).

At the same time, a number of hydrodynamical and magnetohydrodynamical 
simulations were carried out and the theoretical nebulae compared with the 
observed ones (Mellema, 1997; Dwarkadas \& Balick, 1998; Garcia- Segura et 
al., 1999 and references therein). 

Our 3-D reconstruction method, allowing us to derive the detailed
spatial structure of each object and to observe it from all possible 
directions, revolutionizes the approach to the
morphological problem of PNe and opens new prospectives for
understanding the 2-D (morphology) - 3-D (structure) relation.

Let's consider Fig. 6, for instance, giving a representative sample of
the manifold morphologies of NGC~1501 when changing the line of view.

The optical appearance of the nebula mainly depends on the
``latitude'', i. e. on the angle of view of the dense, inhomogeneous
equatorial belt: at ``low latitudes'' (from (0,135) to (0,15) in
Fig. 6) it is a broad disk presenting inner, amorphous
structures. When seen almost pole-on (from (0,60) to (0,105)), a
sharp, oval ring appears, brighter along the minor axis (corresponding
to the projection of the c axis of the ellipsoid).

A quick look through the main imagery catalogues
\citep{acker,schwarz,manchado,gorny}
indicates that NGC 1501-rebuilt resembles M 3-30 when seen from
(0,15), A 53 and NGC 7094 from (0,30), IC 1454 from (0,45), A 73 from
(0,60), A 70 from (0,105), NGC 4071 from (0,120) and NGC 6894 from
(0,135).

We can add that, seen from other directions (not presented here for 
reasons of space), NGC 1501 looks like several other nebulae contained in the 
above-mentioned catalogues.  

In short: all the possible morphologies assumed by NGC 1501 when seen
from different directions fall in the categories named C (centric PN)
by \citet{greig}, Def (elliptical disk with filaments) by \citet{zuck}, 
E (elliptical) by \citet{balick}, ES (elliptical with inner
filaments) by \citet{stan94}, E (elliptical) by Corradi \& Schwarz 
(1995) and
Es (elliptical with inner structure) by \citet{manchado}. They
correspond to quite regular and symmetric shells, rings or disks
probably ejected by single, low mass progenitors (1.1 M$\odot$ $<$
M$_{MS}$ $<$ 2.5 M$\odot$), excited by low mass central stars (0.56
M$\odot$ $<$ M$_{CS}$ $<$ 0.60 M$\odot$) showing a galactic
distribution and kinematics typical of the old Disk Population
(Peimbert \& Serrano, 1980; Corradi \& Schwarz, 1995; Maciel \&
Quireza, 1999 and references therein).

In a certain sense, all this belongs to the past:
the understanding of the different mechanisms and physical processes
forming and shaping PNe passes through the knowledge of the envelope
structures. In their turn, these 3-D structures are usually presumed
from a schematic classification of the manifold nebular
morphologies. Such a rough cascade process is overcome by our 3-D
analysis, which directly provides the accurate spatial distribution
(and ionization) of the expanding gas, thus removing any misleading
camouflage due to projection.

An example: while it is evident that in no case will NGC 1501 appear
as a butterfly or a hourglass, the opposite could occur: when seen at
particular directions (e. g. along -or close to- the axis of the
lobes), a genuine butterfly or hourglass PN (like NGC 2346 or NGC
650-1) could mimic some NGC 1501 appearances (as suggested for K 4-55
by Guerrero et al., 1996).

This could also be the case for some famous and extensively studied PNe, like 
NGC 6720. The 
three-dimensional structure of the Ring Nebula has never been firmly defined, 
spanning from a complex toroid viewed close to the polar axis 
\citep{mink60,louise,reay,balick92,volk,bryce,lame}
to an oblate or prolate spheroidal or ellipsoidal shell 
\citep{atherton,phil,kupf,masson,pascoli90,guerrero},
to a flat ring \citep{hua} and to a hollow cylinder
\citep{proisy,bachiller}.
No doubt the Ring Nebula will be one of our next targets.

\subsection{Zooming in NGC 1501: local filling factor, $\epsilon_l$, and 
density fluctuations in the main shell}

Figs. 4, 5 and 6 clarify that the big holes characterizing the optical
appearance of NGC 1501 represent localized shell regions of lower
density. This is even more evident in Fig.~9, showing the ``swapped
stereoscopic'' nebular structure for $Ne>700\,{\rm cm}^{-3}$ as seen
from (0,0), i.e. from the Earth.

\begin{figure*}
%\resizebox{17.9cm}{!}{\includegraphics*{MS1018f9.eps}}
\caption{Opaque reconstruction of NGC 1501 for Ne$>$700 cm$^{-3}$, as seen 
from (0,-7; left) and (0,7; right), allowing the reader to have a
``swapped stereoscopic vision'' of the nebula from (0,0), i. e. from
the Earth.  To enjoy the 3-D view, cross your eyes till the two dots
in the upper part of the figure perfectly coincide (in this way the
left eye is forced to observe the right image, and viceversa).
``Swapped stereoscopy'' presents a great advantage with respect to the
``direct'' one: the image size can be as large as one wishes, thus
achieving a better resolution than the ``direct method'' (where the
centres of two adjacent images must be closer than the pupils
separation). The drawback of ``swapped stereoscopy'' is the prohibitive 
physical size of a figure containing a large number of images (as in the 
case of Figs.~4 and 5). Therefore, ``direct stereoscopy'' is convenient 
when presenting many pairs.  
Notice that in both methods a
myope has an advantage, since he/she focuses closer.  Caveat for
beginners: when looking at a swapped stereoscopic pair for a long time, a
slight trouble can arise (due to squint). Don't worry, but be careful.
The field and the orientation are as in Fig. 1.}
%\label{figura9}
\end{figure*}

The hole in PA$= 70^\circ$ at 11 arcsec from the central star
corresponds to a ``window'' in the approaching nebular gas; through
this ``window'' we observe the receding, dense material in the oblique
belt representing the equatorial region of NGC 1501. The hole is
clearly visible also in the following tomographic maps: Fig. 3 
 at PA$=80^\circ$, East sector and Fig. 5 of P1 at
PA$=55^\circ$, N-E sector.

Since in the hole $Ne$(receding gas)/$Ne$(approaching gas) $>
1.6$, we have flux(rec. gas)/flux(app. gas) $> 2.5$; i. e. to a first
approximation we can neglect the contribution of the approaching
material and assume that we are observing only the receding one. In
this way we have isolated a well-defined nebular portion, appropriate 
for study in detail at the maximum spatial resolution (TNG or, better,
HST imagery).

Figures 10a, b and c show the flux maps in the $15\times15$ arcsec$^2$
nebular region centred on the hole at 11 arcsec from the central star
in PA$=70^\circ$, as given by our TNG broad-band R frame (Fig. 10a), by
the same frame after a soft Lucy-Richardson restoration (Fig. 10b),
and by HST (Fig. 10c).

These figures stand out intensity fluctuations in the hole by a factor
of 2.0, 2.5 and 3.0 for the TNG, TNG + Lucy-Richardson and HST images,
respectively, indicating a clear correlation with the angular
resolution.

Let's take in the receding portion of the nebular shell observed
through the hole:

\begin{description}
\item[-] the HST value as representative of the small scale flux variations,
\item[-] the $Ne$ radial profile derived from the 3-D reconstruction,
\item[-] the constancy of the electron temperature.
\end{description}

Under these assumptions we obtain fluctuations of the density peaks of
a factor $1.5(\pm0.2)$.

Moreover, using a black-body spectrum for the exciting star and
``normal'' chemical abundances for the expanding gas (see Aller \&
Ckyzak, 1983), we have that:

- the electron density peaks given by the [SII] red doublet are
$1.10 - 1.20$ larger than $Ne(SB)$;

- due to the incomplete recombination of S$^{++}$, the [SII] 6717/6731
\AA\/ density peaks are 10 to 30\% lower than the true $Ne$ values.

This is valid for a wide range of nebular distances ($1.0 < D
{\rm [Kpc]}< 2.0$) and stellar fluxes ($3.0 < \log L_* (L\odot) < 3.6; 4.90
< \log T_* <5.10$).

Very similar results come from the analysis of other ``holes'' present
in NGC 1501, for instance the one at 15 arcsec from the star in
PA$=125^\circ$, corresponding to a ``window'' in the approaching gas,
and the one at 11 arcsec from the star in PA$= -13^\circ$, representing
a ``window'' in the receding material (this last hole is visible in
Fig. 1 and in Fig. 3 at PA$= 170^\circ$, Northern sector).

In spite of the heavy assumptions, the foregoing results would
indicate small-scale density fluctuations (combined with ionization)
as the main cause of the $Ne({\rm [SII]}) - Ne(SB)$ discrepancy
reported in Section 4.

This implies that:
\begin{description}
\item[-]
 the local filling factor in the main shell is
$\epsilon_l=0.50(\pm0.1)$, i. e. lower than the value derived from
($Ne(SB)/Ne{\rm [SII]})^2= 0.66(\pm0.1$);
\item[-]
 our surface brightness electron densities should be increased by a
factor $1.4(\pm0.2)$ to obtain the density peaks of the emitting gas
in the main shell.
\end{description}

\begin{figure}
%\resizebox{\hsize}{!}{\includegraphics*{MS1018f10.eps}}
\caption{Flux distributions in the nebular region (15x15 arcsec$^2$) 
containing the ``hole'' 
at 11 arcsec from the star in P. A.= 70$^o$, as obtained from our TNG
R frame of NGC 1501 (Fig. 10a), from our TNG R frame after the
application of a soft Lucy-Richardson restoration (point spread
function=seeing and ten iterations; Fig. 10b), and from the available
HST imagery (Fig. 10c; we have assembled the four WFPC2 frames taken
by H. Bond).  The hole is the roundish region 8 arcsec in diameter at
the centre of the 15$\times$15 arcsec$^2$ field.  The field stars
(truncated peaks) projected just outside the hole can be used as
identification markers in Fig. 1.}
%\label{figura10}
\end{figure}

It is evident that new insights into the variation on a small scale of the
physical conditions are needed to deepen our understanding of the 
above-mentioned
questions; this mainly concerns the electron temperature, whose
changes affect in different ways the emissivities of recombination and
forbidden lines (a lower $Te$ favouring recombinations; see Peimbert,
1994, for a recent review).

As suggested in P1, an even lower value of $\epsilon_l$ is
expected in the inwards tail of the radial $Ne$ distribution, this
knotty and inhomogeneous region arising from hydrodynamical processes.

Moreover, the application of eq. (1) to the whole nebula gives a
global filling factor $\epsilon<0.25$ (corresponding to the
parameter normally used to estimate the fraction of the total nebular
volume filled by the ionized gas; see Boffi \& Stanghellini, 1994,
and references therein).

The zooming capability implied in our 3-D analysis is a powerful tool
to overcome the existing gap between the potential resolution of the
modern photo-ionization models and the actual resolution when applied
to real nebulae. The first (for example CLOUDY; Ferland et al., 1998)
can be as detailed as one wishes simply using proper input parameters
(stellar flux distribution, nebular dimensions, density law etc.),
whereas their practical application is limited by observational and
reduction procedures and by projection effects:
up to now, all efforts intended to isolate a well-defined nebular
portion (long-slit spectra analysed at different positions, short-slit
echellograms etc.; see Barker, 1991, Perinotto \& Corradi, 1998 and
Hyung et al., 2000), in the best cases give emission line intensities
which are a mixing of both the approaching gas and receding gas
spectra.

The zooming yield is correlated to the spatial resolution of the
rebuilt nebula (i. e. to the spatial and spectral resolutions of the
echellograms); it is quite modest in the present study of NGC 1501,
performed at spatial and spectral resolutions of $1.5 - 2.5$ arcsec and
$22000 - 25000$, respectively.  The output will increase when analysing
the symmetric knots of the quadrupolar PN IC 4634, the ansae and the
FLIERS of NGC~7009 (the ``Saturn'') and the ``wings'' of the butterfly
PN HB 5 (observed with ESO NTT + EMMI at spatial and spectral
resolutions of 1.0 arcsec and 60000, respectively).

\subsection{Radial electron density profile and open questions}

The intersection of the rebuilt-nebula, seen from (0,0), with the
plane perpendicular to the line of sight and crossing the central
star, gives the $Ne$ map shown in Figure 11, representing the radial
electron density profile (for $\epsilon_l=1$) extended to all
position angles.

Fig. 11 shows in detail the most intriguing characteristic of the gas 
distribution in NGC 1501, i. e. the broad inwards tail present in the 
radial density profile, probably arising from hydrodynamical processes 
(Rayleigh-Taylor instabilities and winds interaction; Capriotti, 1973; 
Kahn \& Breitschwerdt, 1990, Garcia-Segura et al., 1999). 

The first process worked for about 1000 yr in the
early evolutionary phases, when the nebula was ionization bounded (up
to R $\simeq$ 0.06 pc).

The second one, being independent of the optical thickness, acted for much
longer (for most of the nebula's life, i. e. 5000 yr) and is probably
still present, as suggested by the large value of $\dot{M}$
($\log\dot{M}= -6.28 M\odot$ yr$^{-1}$) obtained for the WC4 nucleus
of NGC 1501 by
\citet{koesterke} using the standard atmosphere model of
Wolf-Rayet stars.

This simplified scenario is complicated by the peculiarities of the
powering nucleus of NGC 1501 (very hot W-R, rich in C and O, hydrogen
deficient, pulsating), suggesting that the star suffered a thermal
pulse in the late post-AGB evolution. Its path in the HR diagram is
quite uncertain, given the large number of parameters involved: phase
in the nuclear burning cycle at which the star departs from the AGB;
mass of the residual hydrogen-rich stellar envelope; mass-loss rates
during the quiescent hydrogen-burning phase, the high-luminosity phase
following the helium shell flash and the low-luminosity quiescent
helium burning phase; efficiency of convective overshoot in mixing
hydrogen-free material out to the stellar surface; chemical mixing due
to stellar rotation (Sch\"onberner, 1979; Iben, 1984; Bl\"ocker, 1995
and references therein).

Thus, the detailed analysis of the mechanisms and physical processes
forming and shaping the nebula is deferred to a deeper knowledge of
both the star evolutionary parameters and the gas structure.

A final remark concerns the rarity of the radial density profile shown
by NGC 1501: no evidence of an extended inward tail is present in the
other two dozen PNe up to now spectroscopically observed by us with
the Asiago 1.82m telescope + Echelle, ESO NTT + EMMI and TNG + SARG.

A good candidate could be NGC 6751; the complex morphology of this
nebula (excited by a WC4 star) was shown by \citet{corradi} and
spectroscopically studied at high dispersion by \citet{chu91}, who
identified six different components having as many spatio-kinematical
characteristics, but didn't give information on the radial density
profile of the ionized gas.

All this, on the one hand confirms and extends the results already obtained 
in P1, and highlights the potentialities of our reconstruction method, 
on the other hand leaves unresolved many exciting questions, such as:

- what is the small-scale matter distribution in the 
main shell? And that in the inward tail?

- did the hydrogen-depleted stellar wind create 
chemical composition gradients across the nebula?

- how large are the electron temperature and turbulence fluctuations 
in the shell?

- what are the physical conditions, dynamics, ionic and total abundances
in the outer, roundish halo?

- is there any observational evidence supporting the ``born-again scenario''
proposed by Iben et al. (1983; see also Bl\"ocker, 1995, and Herwig et al.,
1999) for nonradial g-mode pulsators?

Most of these (and many other) questions can be answered by a deeper 
spectroscopic study at higher spatial and spectral resolutions (and at 
more position angles).

\begin{figure}
%\resizebox{\hsize}{!}{\includegraphics*{MS1018f11.eps}}
\caption{$Ne$ map and isophotal contours in the sky plane crossing the 
central star when 
the nebula is seen from (0,0), showing the peculiar radial density
profile of NGC 1501. The density contours cover the range 200
cm$^{-3}$ to 1200 cm$^{-3}$ and are spaced by 100 cm$^{-3}$.  This
figure corresponds to the zero-velocity pixel column extended to all
position angles (when using high resolution, slit spectroscopy), and
to the rest-frame (for Fabry-Perot imagery).}
%\label{figura11}
\end{figure} 

\subsection{Application of the 3-D analysis to expanding nebulae in general}

PNe, nova shells, nebulae ejected by symbiotic stars, bubbles
surrounding early spectral type main sequence stars, supernova
remnants (SNRs), shells around Population I W-R stars all
consist of expanding masses of ionized gas, and the many
facets of the 3-D analysis can be successfully applied to them for studying
morphology, dynamics, physical conditions, photoionization model and
evolutionary status.

In all cases the starting point is high resolution slit spectroscopy (or
Fabry-Perot interferometry). Given the large range of expansion
velocities shown by the different classes of nebulae (a few
dozen km s$^{-1}$ for PNe to several hundred km s$^{-1}$ for young
SNRs and nova shells), the term ``high resolution'' is here intended
relative to V$_{exp}$.

Thus, the first parameter to be considered is the ``relative resolution'', 
RR=V$_{exp}/\Delta$V, $\Delta$V being the spectral resolution.

The best RR value is always a compromise between two contrasting
factors, dispersion and detection: a bright nebula can be observed at
large RR in both strong and weak lines, whereas only the brightest
emissions are detectable in a low surface brightness object (even at
small RR).

An indicative lower limit to the applicability of the 3-D analysis can
be put at RR=3. This means that slow expanding nebulae (like PNe) need
high dispersion spectra ($\lambda$/$\Delta$$\lambda $$>$ 20000),
whereas low dispersion spectroscopy ($\lambda$/$\Delta$$\lambda$ $\simeq$
1000) is preferable for the fastest objects. The latter is clearly
illustrated by Lawrence et al. (1995), who derived the three
dimensional model of the emitting gas in the Crab Nebula, in Cassiopea
A and in the remnant of GK Per (Nova Persei 1901) by means of low
resolution (200 $<$ $\lambda$/$\Delta$$\lambda$ $<$ 1500) Fabry-Perot
imagery.

A second parameter to be considered in the 3-D analysis is the spatial
resolution along the slit ($\Delta$d).  Also in this case:

- it is intended ``relative to the angular dimensions $d$ of the nebula'';

- the larger the ``relative spatial resolution'' $SS=d/\Delta d$, the
better the reconstruction;

- as a rule of thumb, SS must be larger than 3. 

A detailed discussion of the various contributions given by the 3-D
analysis in solving the manifold problems connected with the different
classes of expanding nebulae is beyond the aims of this paper and is
left to the education, experience and imagination of the reader.  In
fact, our intent is exactly this: to excite the interest of some other
people, exposing them to tackle the many problems and the wide
applications of the 3-D methodology (we feel quite alone at
present... ``rari nantes in gurgite vasto'' paraphrasing Virgil,
Aeneid I, 118).

\section{Conclusions}

This paper introduces a general reduction procedure giving the detailed
spatial structure of the ionized gas in expanding nebulae.

Its application to NGC 1501 indicates that this high excitation PN is an
almost oblate ellipsoid of moderate ellipticity ($a \simeq 44$ arcsec, $a/b
\simeq 1.02$, $a/c \simeq 1.11$), denser in the equatorial belt containing 
the axes $b$ and $c$.  A lumpy, quite homogeneous cocoon completely
embeds the main shell.

A series of images is presented, showing stereoscopic views of the 3-D 
gas structure and the nebular appearance when changing the line of 
sight.

The morphology-structure correlation is discussed, the small scale
density variations in the main shell and the peculiarities of the
matter radial profile are analysed.

NGC 1501 is a quite faint, high ionization, density bounded PN, its
3-D study is simple and straightforward (at our spatial and spectral
resolutions).

We are now involved in a much more ambitious project: the spatial
mapping in different ions (H$^+$, He$^+$, He$^{++}$, O$^o$, O$^+$,
O$^{++}$, N$^+$, S$^+$, Ne$^{++}$, Ar$^{++}$, Ar$^{+++}$ etc.) and the
photo-ionization modelling of a dozen bright PNe in both hemispheres;
each target observed at several (at least 9) position angles with 
ESO NTT + EMMI (spatial resolution = 1.0 arcsec,
spectral resolution = 60000) and TNG + SARG (spatial resolution = 0.7
arcsec, spectral resolution = 115000).

\begin{acknowledgements}

We greatly appreciated the suggestions and the encouragement by Luciana Bianchi, 
Gary Ferland, Guillermo Garcia-Segura and Vincent Icke; 
to them all we express our deep gratitude.

This paper is based on observations made with:

- the 1.82m telescope of the Astronomical Observatory of Padua, operated at 
Asiago, Cima Ekar (Italy);

- the Italian Telescopio Nazionale Galileo (TNG) operated on the
island of La Palma by the Centro Galileo Galilei of the CNAA
(Consorzio Nazionale per l'Astronomia e l'Astrofisica) at the Spanish
Observatorio del Roque de los Muchachos of the Instituto de
Astrofisica de Canarias;

- the NASA/ESA Hubble Space Telescope, obtained from the data archive
at the Space Telescope Science Institute. STScI is operated by the
Association of Universities for Research in Astronomy, Inc. under NASA
contract NAS 5-26555.

\end{acknowledgements}

\bibliographystyle{apj}
%\bibliography{}

\end{document}